\documentclass[a4paper,12pt]{article}

\begin{document}
\topmargin 0pt \oddsidemargin 0mm

\renewcommand{\thefootnote}{\fnsymbol{footnote}}
\begin{titlepage}

\vspace{2mm}
\begin{center}
{\Large \bf Hawking Radiation as Quantum Tunneling in Rindler
Coordinate}
 \vspace{12mm}

{\large Sang Pyo Kim\footnote{e-mail address: sangkim@kunsan.ac.kr}}\\
 \vspace{5mm}
{\em Department of Physics, Kunsan National University, Kunsan
573-701,
 Korea \\Asia Pacific Center for Theoretical Physics, Pohang 790-784,
 Korea}\\

\end{center}

\vspace{45mm} \centerline{{\bf{Abstract}}} \vspace{5mm} We
substantiate the Hawking radiation as quantum tunneling of fields or
particles crossing the horizon by using the Rindler coordinate. The
thermal spectrum detected by an accelerated particle is interpreted
as quantum tunneling in the Rindler spacetime. Representing the
spacetime near the horizon locally as a Rindler spacetime, we find
the emission rate by tunneling, which is expressed as a contour
integral and gives the correct Boltzmann factor. We apply the method
to non-extremal black holes such as a Schwarzschild black hole, a
non-extremal Reissner-Nordstr\"{o}m black hole, a charged Kerr black
hole, de Sitter space, and a Schwarzschild-anti de Sitter black
hole.
\\

\end{titlepage}

\newpage
\renewcommand{\thefootnote}{\arabic{footnote}}
\setcounter{footnote}{0} \setcounter{page}{2}

\section{Introduction}

A black hole radiates thermal radiation with the Hawking temperature
determined by the surface gravity at the event horizon
\cite{Hawking}. The surface gravity is the acceleration measured at
the spatial infinity that a stationary particle should undergo to
withstand the gravity at the event horizon. The accelerated particle
detects a thermal spectrum with the Unruh temperature out of the
Minkowski vacuum \cite{Unruh}. The particle accelerated with the
surface gravity would see the vacuum containing a thermal flux with
the Hawking temperature. The thermal spectrum seen by the
accelerated particle can also be understood by the interpretation
that the Minkowski vacuum is restricted to a causally connected
Rindler wedge due to presence of horizons just as the horizon of a
black hole prevents the outer region from being causally connected
with the inner horizon \cite{Israel}.

Recently Parikh and Wilczek reinterpreted the Hawking radiation as
quantum tunneling \cite{Parikh}. Their observation is that as a
particle has a negative energy just inside and a positive energy
just outside the horizon, a virtual pair created near the horizon
can materialize into a real pair with zero total energy, one
particle on each side of the horizon. Indeed, the Hawking radiation
is such particle production and can be interpreted as tunneling
through the horizon. The tunneling process for particle production
by geometry of a black hole is analogous to the Schwinger mechanism
for pair production by an external electric field \cite{Schwinger}.
Indeed, charged pairs can be materialized from virtual pairs when
the potential energy across the Compton wavelength is comparable to
the rest mass of the particle.

One approach to the tunneling interpretation of Hawking radiation is
to study the tunneling motion of the $s$-wave of emitted radiations,
which has been applied to various black holes
\cite{Kraus95-1}-\cite{Kerner07} (for a review, see
\cite{Padmanabhan}). Another approach is to study quantum fields
tunneling through the horizon in a black hole spacetime
\cite{Srinivasan}. In the latter a conventional wisdom is to find
the imaginary part of the Hamilton-Jacobi action, twice of which
leads to a Boltzmann factor \cite{Srinivasan}-\cite{Kim07}. However,
the emission rate by tunneling depends on the coordinate system. The
radial coordinate, for instance, yields a temperature twice of the
Hawking temperature. Some proposals were advanced to remedy the
ambiguity. In ref. \cite{Srinivasan}, the ratio of emission to
absorption was used to get the correct Boltzmann factor. Also the
isotropic coordinate and the proper distance from the horizon was
used \cite{Angheben}, \cite{Nadalini}, and different coordinates
were tested \cite{Shankaranarayanan02}. In refs. \cite{Chowdhury}
and \cite{Akhmedov} the emission rate was suggested as a contour
integral, which makes the rate invariant under canonical
transformations.

The purpose of this paper is to substantiate the tunneling idea
within the context of quantum field theory in the black hole
spacetime. The analogy between the Hawking radiation and the
Schwinger mechanism has been well exploited
\cite{Brout}-\cite{Parentani97}. As the Schwinger mechanism can be
interpreted as the Unruh effect seen by a charged particle
accelerated by the electric field, we may use the Rindler coordinate
for the accelerated particle. In addition, as the Hawking radiation
can be interpreted as the Unruh effect with the surface gravity, the
Rindler coordinate may be used to describe the tunneling process of
quantum fields. In the tunneling interpretation a local Rindler
frame was first introduced in ref. \cite{Padmanabhan04}, the Unruh
effect was calculated from tunneling \cite{Akhmedov07}, and the
discrepancy of the Hawking temperature from tunneling from the left
to the right wedge and from the right to the future wedge in the
full Rindler spacetime was discussed \cite{Nakamura}. In the
previous paper \cite{Kim07}, along the lines of the Schwinger
mechanism, the Rindler coordinate was used to get the correct
Boltzmann factor for a charged black hole and a BTZ black hole.

Another reason for using the Rindler coordinate is that quantum
tunneling of fields or particles occurs through the horizon and
locally the Rindler coordinate near the horizon is an accelerated
frame for the Minkowski spacetime when the acceleration is the
surface gravity. Further the field equation in the Rindler spacetime
has many properties in common with the equation minimally coupled
with the gauge field of electric field for the Schwinger mechanism.
All these imply that the Rindler coordinate is an appropriate
coordinate to make full use of the analogy between the Schwinger
mechanism and the Hawking radiation. We show that in the Rindler
spacetime the tunneling rate of a quantum field from a causally
disconnected region through the horizon explains indeed the thermal
spectrum with the Unruh temperature. In fact, the tunneling rate in
the Rindler spacetime takes the same form as the rate for tunneling
of virtual pairs in the Schwinger mechanism.

We represent a black hole spacetime in the Rindler coordinate and
then calculate the emission rate by tunneling of quantum fields
through the horizon. The emission rate found in analogy with the
Schwinger mechanism is given by a contour integral in the Rindler
spacetime. As the contour integral is independent of coordinate
transformations, the emission rate is invariant under canonical
transformations. Also as quantum physics just inside and outside
horizons can be properly described by Rindler coordinates, the
Rindler spacetime seems to be a natural coordinate that avoids the
ambiguity from the modification of contours in changing coordinates.
We get the correct Boltzmann factor for non-extremal black holes
such as a Schwarzschild black hole, a non-extremal
Reissner-Nordstr\"{o}m black hole, a charged Kerr black hole, a de
Sitter space, and a Schwarzschild-anti de Sitter black hole.
However, our method cannot be applied to extremal black holes
because there are no Rindler coordinates locally near horizons.

The organization of this paper is as follows. In section 2 we
discuss some ambiguity of the Hawking radiation as tunneling. In
section 3 we derive the emission rate by tunneling in the Rindler
spacetime and then compare it with the Schwinger mechanism. In
section 4 we use the Rindler coordinate to derive the emission rate
within quantum field theory and interpret the Hawking radiation as
quantum tunneling for a Schwarzschild black hole, a non-extremal
Reissner-Nordstr\"{o}m black hole, a charged Kerr black hole, a de
Sitter space, and a Schwarzschild-anti de Sitter black hole. In
section 5 we compare the emission rate in the Rindler coordinate
with the isotropic coordinate and the proper distance from the
horizon. Finally we conclude in section 6.

\section{Hawking Radiation as Tunneling}

In this section we briefly review tunneling of quantum fields
through the horizon in a black hole spacetime and discuss the
related problems. It was pointed out that this approach to the
tunneling interpretation of Hawking radiation has ambiguities such
coordinate-dependence of the Hawking temperature and non-invariance
of the action under canonical transformations \cite{Chowdhury},
\cite{Akhmedov}.

For the sake of simplicity we consider a stationary black hole with
the metric of the form
\begin{eqnarray}
ds^2 = - f(r) dt^2 + \frac{dr^2}{g(r)} + h_{ij} dx^i dx^j,
\label{metric}
\end{eqnarray}
where $h_{ij} dx^i dx^j$ is a two-dimensional metric and, for
instance, becomes $r^2 d \Omega^2$ for a spherically symmetric black
hole. The event horizon $r_H$ is located at  $f (r_H) = g(r_H) = 0$,
near which  $f (r) = f'(r_H) (r - r_H) $ and $g (r) = g' (r_H) (r -
r_H)$ up to the leading term. The exceptional case of extremal black
holes will be  treated separately. As we are mostly concerned about
the $s$-wave (spherically symmetric) of a massive scalar field in
the black hole spacetime, we shall further restrict our
investigation to the two-dimensional sector of $(t,r)$.

The equation of the massive scalar takes the form [in units with
$\hbar = c = 1$]
\begin{eqnarray}
\Biggl[ \frac{1}{\sqrt{fg}} \frac{\partial^2}{\partial t^2} -
\frac{\partial}{\partial r} \Bigl(\sqrt{fg} \frac{\partial}{\partial
r} \Bigr) + m^2 \Biggr] \Phi (t, r) = 0.
\end{eqnarray}
The action from the solution, $\Phi (t, r) = e^{i S_{\pm} (t,r)}$,
with
\begin{eqnarray}
S_{\pm} (t, r) = - \omega t \pm \int p(r) dr,
\end{eqnarray}
satisfies, at the leading order, the Hamilton-Jacobi equation
\begin{eqnarray}
\Biggl(\frac{\partial S_{\pm}}{\partial r} \Biggr)^2 -
\frac{\omega^2}{fg} + \frac{m^2}{\sqrt{fg}} = 0.
\end{eqnarray}
Here $ \pm p (r) = \partial S_{\pm}/\partial r$ is the radial
momentum of an outgoing or ingoing wave, respectively. The action is
then given by
\begin{eqnarray}
S_{\pm} (t,r) = - \omega t \pm \int_{r_0}^r \frac{dr}{\sqrt{fg}}
\sqrt{\omega^2 - m^2 \sqrt{fg}}. \label{phase}
\end{eqnarray}
There is an ultraviolet divergence from the simple pole at the event
horizon, since $f g = f'(r_H) g'(r_H) (r - r_H)^2$, which
contributes to an imaginary part.

To describe waves tunneling through the event horizon, $r_0$ is
located inside the horizon. The imaginary part may be obtained by
taking a semi-circle under the horizon $r_H $ as
\begin{eqnarray}
{\rm Im} S_{\pm} = \pm \frac{\pi  \omega}{\sqrt{f' (r_H) g'(r_H)}}.
\end{eqnarray}
Then the amplitude square, $|\Phi|^2 = e^{- 2 {\rm Im} S_+}$, leads
to the tunneling (emission) rate for the outgoing wave through the
horizon:
\begin{eqnarray}
P = e^{- 2 {\rm Im} S_+} = e^{- \omega /T},
\end{eqnarray}
where
\begin{eqnarray}
T = \frac{\sqrt{f' (r_H) g'(r_H)}}{2 \pi}. \label{tem}
\end{eqnarray}

A few remarks are in order. First, note that the temperature
(\ref{tem}) is the twice of the Hawking temperature
\begin{eqnarray}
T_H = \frac{\sqrt{f' (r_H) g'(r_H)}}{4 \pi}.
\end{eqnarray}
Several proposals were advanced to remedy the ambiguity of the
temperature. In ref. \cite{Srinivasan},  by analogy with a black
body  the emission rate is defined as the ratio of the amplitude
square of the outgoing wave to that of the ingoing wave, which is
given by $S_-$,
\begin{eqnarray}
\frac{P_{out}}{P_{in}} = e^{- 2 {\rm Im} (S_+ - S_-)}   = e^{-
\omega / T_{H}}.
\end{eqnarray}
On the other hand, the isotropic coordinate and the proper distance
along the radial direction are also employed to get the correct
Hawking temperature in refs. \cite{Angheben}, \cite{Nadalini}. In
fact, the ambiguity of the temperature is originated from the
coordinate used to calculate the emission rate. Another ambiguity is
that the action (\ref{phase}), in particular, the imaginary part is
not invariant under canonical transformations \cite{Chowdhury},
\cite{Akhmedov}.

\section{Tunneling in the Rindler Spacetime}

A Rindler spacetime is the spacetime covered by all time-like
congruences of an accelerated particle. The Rindler spacetime has
two horizons that separate the Minkowski spacetime into the right
(R), the left (L), the past (P) and the future (F) wedges. The wedge
(R) is causally disconnected from the wedge (L) and a timelike
Killing vector normal to the spacelike Cauchy surface that covers
both (R) and (L) defines a complete set of quantum fields. The
thermal nature of Unruh effect comes from the fact that the
physically accessible region (R) for an accelerated particle is
causally disconnected with the other region (L) and the Minkowski
vacuum defined in the union of (R) and (L) looks like a mixed state
for the particle \cite{Israel}. In this paper, following the
arguments in refs. \cite{Unruh} and \cite{Israel}, we shall consider
quantum tunneling from (L) to (R) wedge. However, in ref.
\cite{Nakamura} it was shown that the result of tunneling from (R)
to (F) would differ from the Hawking temperature by a factor of two.

In two dimensions the right wedge (R) of the Rindler spacetime has
the coordinate
\begin{eqnarray}
t &=& \rho_R \sinh (a \tau), \nonumber\\
z &=& \rho_R \cosh (a \tau), \label{r wedge}
\end{eqnarray}
with $\rho_R \geq 0$, and the left wedge (L) has
\begin{eqnarray}
t &=& \rho_L \sinh (a \tau), \nonumber\\
z &=& \rho_L \cosh (a \tau), \label{r wedge}
\end{eqnarray}
with $\rho_L \leq 0$. Here $a$ is the acceleration of the particle.
In both wedges the spacetime has the metric
\begin{eqnarray}
ds^2 = - (a \rho)^2 d \tau^2 + d \rho^2.
\end{eqnarray}
The right wedge is causally disconnected from the left wedge by
horizons, $t = \pm z$, which correspond to $\rho  = 0$. The
accelerated particle would detect a thermal spectrum with the
so-called Unruh temperature, $T_U = a/ (2 \pi)$, from the Minkowski
vacuum \cite{Unruh}. From a view point of causality, the particle
operators for the accelerated particle in (R) are expressed in terms
of Minkowski operators in (R) and (L) through a Bogoliubov
transformation, so the vacuum for the particle looks like a thermal
vacuum in thermo-field dynamics \cite{Israel}.

However, quantum mechanically speaking, fields  or particles can
cross horizons with a certain probability. To cover both (R) and
(L), we analytically continue the coordinate
\begin{eqnarray}
\rho_L = \rho_R e^{i \pi}. \label{anal con}
\end{eqnarray}
The massive scalar field in (R) and (L) obeys the equation
\begin{eqnarray}
\Biggl[- \frac{1}{(a \rho)^2} \frac{\partial^2}{\partial \tau^2} +
\frac{\partial^2}{\partial \rho^2} - m^2 \Biggr] \Phi (\tau, \rho) =
0, \label{rin eq1}
\end{eqnarray}
and the spatial part, $\Phi = e^{- i \omega \tau} \varphi (\rho)$,
satisfies the equation
\begin{eqnarray}
\Biggl[\frac{\partial^2}{\partial \rho^2} + \frac{\omega^2}{(a
\rho)^2} - m^2 \Biggr] \varphi (\rho) = 0. \label{rin eq2}
\end{eqnarray}
Note that eq. (\ref{rin eq2}) is oscillatory for $|\rho| \leq \rho_c
= \omega/(ma)$ and exponential otherwise. In quantum mechanics, it
is a one-dimensional problem with the energy $-m^2$ and with the
negative singular potential $- (\omega/a \rho)^2$.

For a tunneling wave function crossing $\rho = 0$, we may use the
solution in (L)
\begin{eqnarray}
\varphi_L (\rho) = \sqrt{\rho} J_{i \nu} (i m \rho), \label{rin sol}
\end{eqnarray}
where $J_{i \nu}$ is the Bessel function with a complex order
\begin{eqnarray}
\nu = \sqrt{\frac{\omega^2}{a^2}- \frac{1}{4}}.
\end{eqnarray}
The wave function in (R) tunneled from (L) may be found by
analytically continuing the solution (\ref{rin sol}) via (\ref{anal
con}), which becomes
\begin{eqnarray}
\varphi_R (\rho) &=& (\rho_R e^{i \pi})^{1/2} J_{i \nu} (i m \rho_R
e^{i \pi}) \nonumber\\
&=& e^{- \nu \pi} e^{i \pi/2} (\rho_R)^{1/2} J_{i \nu} (i m \rho_R),
\label{r wedge}
\end{eqnarray}
where the relation $J_{\alpha} (z e^{i n \pi}) = e^{i n \alpha \pi}
J_{\nu} (z)$ for an integer $n$ is used. Indeed, the solution
(\ref{r wedge}) has an outgoing flux near the horizon in (R).
Therefore we find the tunneling (emission) rate as the ratio of the
amplitude square of the tunneled wave function in (R) from (L) to
the amplitude square of the outgoing wave function $(\rho_R)^{1/2}
J_{i \nu} (i m \rho_R)$ with a given flux in (R):
\begin{eqnarray}
P = e^{- 2 \nu \pi} \approx e^{- 2 \pi \omega /a}.
\end{eqnarray}
The tunneling rate is nothing but the Boltzmann factor with the
Unruh temperature $T_U$. We further show that the tunneling rate, $P
= e^{- {\cal S}_{\omega}}$, can also be obtained from the action,
$\varphi = e^{i S (\rho)}$, where
\begin{eqnarray}
{\cal S}_{\omega} = 2~{\rm Im} S = - i \oint
\sqrt{\frac{\omega^2}{(a \rho)^2} - m^2} d \rho = \frac{2 \pi
\omega}{a}, \label{rin inst}
\end{eqnarray}
with a contour enclosing $\rho = 0$.

We now compare the tunneling (emission) rate (\ref{rin inst}) in the
Rindler coordinate with the pair production rate by an electric
field in the Schwinger mechanism. In two dimensions the scalar field
equation for charge $e$ ($e > 0$) and mass $m$  minimally coupled
with the Coulomb gauge, $A_{\mu} = (-Ex, 0)$, takes the form
\begin{eqnarray}
 \Biggl[ \Bigl(\frac{\partial}{\partial t} - i e E x \Bigr)^2
 - \frac{\partial^2}{\partial x^2} + m^2 \Biggr] \Phi = 0.
\label{kg eq}
\end{eqnarray}
The spatial part, $\Phi = e^{- i \omega t} \varphi (x)$, satisfies
\begin{eqnarray}
 \Biggl[ \frac{\partial^2}{\partial x^2}
 + (\omega +  e E x )^2 - m^2 \Biggr] \varphi (x) = 0. \label{kg sp}
\end{eqnarray}
In quantum mechanics, eq. (\ref{kg sp}) is a tunneling problem with
energy $-m^2$ under the inverted harmonic potential. In refs.
\cite{Kim-Page07}, \cite{Kim-Page02}, \cite{Kim-Page06}, the
tunneling rate
\begin{eqnarray}
P = e^{- {\cal S}}
\end{eqnarray}
is given by the WKB instanton action
\begin{eqnarray}
{\cal S}  = - i \oint \sqrt{(\omega + eEx)^2 - m^2} dx = \frac{\pi
m^2}{eE}. \label{action}
\end{eqnarray}
Here the contour integral is taken outside a contour in the complex
$x$-plane.

We notice a similarity that the tunneling rate is given by the same
formula
\begin{eqnarray}
P = e^{- i \oint p},
\end{eqnarray}
where for the Rindler case $p$ is
\begin{eqnarray}
p (\rho) = \sqrt{\frac{\omega^2}{(a \rho)^2} - m^2},
\end{eqnarray}
while for the Schwinger mechanism
\begin{eqnarray}
p (x)  = \sqrt{(\omega + eEx)^2 - m^2}.
\end{eqnarray}
The contours include segments of real axis where $p$ is real.
However, for the Schwinger mechanism the contour integral is taken
outside a contour that excludes a branch cut connecting two roots,
$x_{\pm} = (- \omega \pm m)/(eE)$, while for the Rindler case the
contour is taken inside a contour that excludes branch cuts from a
root $\omega/(am)$ to the positive infinity and from another root $-
\omega/(am)$ to the negative infinity.

\section{Tunneling Rate of Black Holes in the Rindler Coordinate}

From the argument in section 3 we shall use the Rindler coordinate
of a black hole spacetime to calculate the tunneling rate. The
spacetime region of the metric (\ref{metric}) near the event horizon
may be locally approximated by a Rindler spacetime
\begin{eqnarray}
ds^2 = - (\kappa \rho)^2 dt^2 + d \rho^2, \label{rindler}
\end{eqnarray}
where the Rindler coordinate is used
\begin{eqnarray}
\kappa \rho = \sqrt{f}, \quad \frac{dr}{d\rho} = \sqrt{g}.
\label{rin eq}
\end{eqnarray}
From eq. (\ref{rin eq}) we find the surface gravity
\begin{eqnarray}
\kappa = \frac{f'(r_H)}{2} \sqrt{\frac{g(r_H)}{f(r_H)}} =
\frac{\sqrt{f'(r_H) g'(r_H)}}{2}. \label{surface gr}
\end{eqnarray}
The last equality holds only for non-extremal black holes. In the
Rindler coordinate the instanton action is the contour integral in
the complex $\rho$-plane
\begin{eqnarray}
{\cal S}_{\omega} = - i \oint \sqrt{\frac{\omega^2}{f(\rho)} - m^2}
d\rho = \frac{2 \pi \omega}{\kappa}. \label{bh action2}
\end{eqnarray}
Here the contribution comes from the simple pole at $\rho = 0$,  the
event horizon. This leads to the emission rate for the Hawking
radiation
\begin{eqnarray}
P (\omega) = e^{- {\cal S}_{\omega}} = e^{ - \omega /T_H}, \quad T_H = \frac{\kappa}{2 \pi}.
\label{tunneling}
\end{eqnarray}
As a by-product the contour integral is invariant under canonical
transformations \cite{Akhmedov}.

In the below we apply the emission rate in the Rindler coordinate to
a Schwarzschild black hole, a non-extremal Reissner-Nordstr\"{o}m
black hole, a charged Kerr black hole, a de Sitter space and a
Schwarzschild-anti de Sitter black hole.

\subsection{Schwarzschild Black Hole}

The Schwarzschild black hole with
\begin{eqnarray}
f = g = 1 - \frac{2M}{r} = \frac{r - r_H}{r}, \quad (r_H = 2M)
\end{eqnarray}
has the metric
\begin{eqnarray}
ds^2 = - (\kappa \rho)^2 dt^2 + \frac{(2 r_H \kappa)^2}{(1 - (\kappa
\rho)^2)^4} d \rho^2, \label{sch bh}
\end{eqnarray}
in the Rindler coordinate
\begin{eqnarray}
f = g = (\kappa \rho)^2.
\end{eqnarray}
Near the event horizon, $\rho \approx 0$ $(r \approx r_H)$, the
spacetime (\ref{sch bh}) approximately becomes a Rindler one with
$\kappa = 1/(2 r_H) = 1/ (4M)$.  Further, if the Euclidean time
$\tau = it$ has a periodicity of $\tau =  2 \pi/ \kappa$, it is the
Euclidean space without a deficit angle. In fact, $\kappa$ is the
surface gravity
\begin{eqnarray}
\kappa = \frac{f'(r_H)}{2} = \frac{1}{4M},
\end{eqnarray}
and leads to the emission rate (\ref{tunneling}) with the Hawking
temperature
\begin{eqnarray}
T_H = \frac{1}{8 \pi M}.
\end{eqnarray}

\subsection{Reissner-Nordstr\"{o}m Black Hole}

The non-extremal Reissner-Nordstr\"{o}m black hole with
\begin{eqnarray}
f = g = 1 - \frac{2M}{r} + \frac{Q^2}{r^2} = \frac{(r- r_+)(r-
r_-)}{r^2},
\end{eqnarray}
has the event horizon, $r_+ = M + \sqrt{M^2 - Q^2}$, and the inner
horizon, $r_- = M - \sqrt{M^2 - Q^2}$. With the coordinate
\begin{eqnarray}
f = g = \frac{r_+^2}{r^2} (\kappa \rho)^2,
\end{eqnarray}
the metric becomes
\begin{eqnarray}
ds^2 &=& - \frac{(2 r_+)^2}{(r_+ + r_- + \sqrt{(r_+ - r_-)^2 + 4
(\kappa r_+ \rho)^2})^2} (\kappa \rho)^2 dt^2 \nonumber\\&& +
\frac{(\kappa r_+)^2 (r_+ + r_- + \sqrt{(r_+ - r_-)^2 + 4 (\kappa
r_+ \rho)^2})^2}{(r_+ - r_-)^2 + 4 (\kappa r_+ \rho)^2} d \rho^2.
\label{rn bh}
\end{eqnarray}
Near the event horizon, $\rho \approx 0$, the spacetime (\ref{rn
bh}) approximately becomes a Rindler one provided that $\kappa  =
(r_+ - r_-)/(2 r_+^2)$, which is the surface gravity at the event
horizon $r_+$,
\begin{eqnarray}
\kappa = \frac{f'(r_+)}{2} = \frac{r_+ - r_-}{2 r_+^2}.
\end{eqnarray}
The emission rate (\ref{tunneling}) is thus valid for non-extremal
Reissner-Nordstr\"{o}m black holes.

A caveat is that our emission rate relies on the Rindler coordinate.
However, the extremal Reissner-Nordstr\"{o}m black hole with $Q = M$
\begin{eqnarray}
ds^2 = - \Biggl(1 - \frac{M}{r} \Biggr)^2 dt^2 +
\frac{dr^2}{\Biggl(1 - \frac{M}{r} \Biggr)^2}
\end{eqnarray}
cannot be approximated by the Rindler spacetime, since with $r = M +
c\rho$ for any $c$, the metric has the form
\begin{eqnarray}
ds^2 = - \frac{(c \rho)^2}{(M+ c \rho)^2} dt^2 + \frac{(M+ c
\rho)^2}{\rho^2} d\rho^2.
\end{eqnarray}
Without the Rindler coordinate the instanton action (\ref{bh
action2}) cannot be applied to extremal black holes.

\subsection{Charged Kerr Black Hole}

The charged Kerr black hole has the metric
\begin{eqnarray}
ds^2 = - f dt^2 + \frac{dr^2}{g} + k (d \phi - \omega dt )^2 +
\Sigma d \theta^2,
\end{eqnarray}
where
\begin{eqnarray}
f &=& \frac{\Delta \Sigma}{(r^2 + a^2)^2 - \Delta a^2 \sin^2
\theta}, \nonumber\\
g &=& \frac{\Delta}{\Sigma}, \nonumber\\
k &=& \frac{(r^2 + a^2)^2 - \Delta a^2 \sin^2 \theta}{\Sigma},
\nonumber\\
\omega &=& \frac{a \sin^2 \theta (r^2 + a^2 - \Delta)}{(r^2 + a^2)^2
- \Delta a^2 \sin^2 \theta},
\end{eqnarray}
where
\begin{eqnarray}
\Delta &=& r^2 - 2 Mr + a^2 + Q^2, \nonumber\\
\Sigma &=& r^2 + a^2 \cos^2 \theta.
\end{eqnarray}
The event horizon is located at $r_+ = M + \sqrt{M^2 - a^2 - Q^2}$
and the inner horizon at $r_- = M - \sqrt{M^2 - a^2 - Q^2}$. With
the coordinate
\begin{eqnarray}
f = (\kappa x)^2,
\end{eqnarray}
the metric near the event horizon approximately takes the form
\begin{eqnarray}
ds^2 = - (\kappa \rho)^2 dt^2 + \frac{4 \kappa^2 (r_+^2 +
a^2)^2}{(r_+ - r_-)^2} d \rho^2.
\end{eqnarray}
With the surface gravity
\begin{eqnarray}
\kappa = \frac{r_+ - r_-}{2(r_+^2 +a^2)} = \frac{\sqrt{f'(r_+) g'
(r_+)}}{2},
\end{eqnarray}
the charged Kerr black hole can be written in the Rindler coordinate
near the event horizon and has the emission rate (\ref{tunneling}).

\subsection{de Sitter Space}

The de Sitter spacetime with
\begin{eqnarray}
f = g = 1 - \frac{r^2}{l^2} = (\kappa \rho)^2
\end{eqnarray}
has the event horizon at $r_H = l$. In the Rindler coordinate the de
Sitter space becomes
\begin{eqnarray}
ds^2 = - (\kappa \rho)^2 dt^2 + \frac{(l \kappa)^2}{1 - (\kappa
\rho)^2} d \rho^2,
\end{eqnarray}
and locally a Rindler spacetime when $\kappa = 1/l$, which is the
surface gravity. Thus the emission rate (\ref{tunneling}) is also
valid for the de Sitter with the Hawking temperature
\begin{eqnarray}
T_H = \frac{1}{2 \pi l}.
\end{eqnarray}

\subsection{Schwarzschild-anti de Sitter Black Hole}

The Schwarzschild-anti de Sitter black hole with
\begin{eqnarray}
f = g = 1 - \frac{2M}{r} + \frac{r^2}{l^2}
\end{eqnarray}
has the event horizon at
\begin{eqnarray}
r_H = l \Biggl[ \Biggl(\frac{M}{l} + \sqrt{\frac{1}{27} + (M/l)^2}
\Biggr)^{1/3} + \Biggl(\frac{M}{l} - \sqrt{\frac{1}{27} + (M/l)^2}
\Biggr)^{1/3}\Biggr].
\end{eqnarray}
Near the event horizon, choosing the coordinate
\begin{eqnarray}
f = g \approx (r - r_H) \times 2 \Biggl(\frac{  (r_H/l)^2 +
Ml/r_H^2}{r_H} \Biggr) = (\kappa \rho)^2
\end{eqnarray}
we may write the metric approximately as
\begin{eqnarray}
ds^2 \approx - (\kappa \rho)^2 + \frac{ (l \kappa)^2}{(r_H/2 +
Ml/r_H^2)^2} d \rho^2.
\end{eqnarray}
With the surface gravity
\begin{eqnarray}
\kappa = \frac{1}{l} \Biggr(\frac{r_H}{l} +  \frac{Ml}{r_H^2}
\Biggr),
\end{eqnarray}
the metric becomes a Rindler spacetime. Thus the emission rate
(\ref{tunneling}) is also valid for the Schwarzschild-anti de Sitter
black hole.

\section{Connection with Other Coordinates}

In this section we discuss why the isotropic coordinate and the
proper distance can recover the correct Hawking temperature and the
Boltzmann factor \cite{Angheben}, \cite{Nadalini}. The spherically
symmetric metric in eq. (\ref{metric}) can be written in the
isotropic coordinate as
\begin{eqnarray}
ds^2 = - f (\zeta) dt^2 + k (\zeta) (d \zeta^2 + \zeta^2 d
\Omega^2),
\end{eqnarray}
where
\begin{eqnarray}
\int \frac{d \zeta}{\zeta} = \int \frac{dr}{r \sqrt{g(r)}}.
\end{eqnarray}
For instance, the Schwarzschild black hole has the isotropic
coordinate
\begin{eqnarray}
ds^2 = - \Biggl(\frac{1 - 2M/\zeta}{1 + 2M /\zeta} \Biggr)^2 dt^2 +
\Biggl(\frac{1 + 2M/\zeta}{2} \Biggr)^4 (d\zeta^2 + \zeta^2 d
\Omega^2), \label{iso}
\end{eqnarray}
where $r = \zeta(1 + 2M/ \zeta )^2/4$. The event horizon is located
at $\zeta_H = r_H = 2M$. The metric near the event horizon is
approximately given by
\begin{eqnarray}
ds^2 \approx - \frac{(\zeta - 2M)^2}{(4M)^2} dt^2 + (d\zeta^2 +
\zeta^2 d \Omega^2).
\end{eqnarray}
Setting $ (\zeta - 2M)/4M =  \kappa \rho$ and $\kappa = 1 / 4M$, the
Schwarzschild black hole metric becomes a Rindler one. This is the
reason why the isotropic coordinate leads to the correct result.

Another coordinate is the proper distance
\begin{eqnarray}
\sigma = \int \frac{dr}{\sqrt{g}},
\end{eqnarray}
and its metric metric
\begin{eqnarray}
ds^2 = - f (\sigma) dt^2 + d \sigma^2.
\end{eqnarray}
For non-extremal black holes, the leading terms are
\begin{eqnarray}
f = f'(r_H) (r - r_H), \quad g = g' (r_H) (r - r_H).
\end{eqnarray}
Now the proper distance from the event horizon
\begin{eqnarray}
\sigma = \frac{2}{\sqrt{g' (r_H)}} \sqrt{r - r_H},
\end{eqnarray}
leads to
\begin{eqnarray}
f = f' (r_H) \Biggl(\frac{\sqrt{g'(r_H)}}{2} \Biggr)^2 \sigma^2 =
(\kappa \sigma)^2,
\end{eqnarray}
with the surface gravity (\ref{surface gr}). Therefore the proper
distance method gives the same result as the Rindler coordinate.

\section{Conclusion}

We have studied the tunneling solution of a scalar field in the
Rindler coordinate of a black hole spacetime. The tunneling solution
in the black hole coordinate leads to a Boltzmann factor with a
temperature twice of the Hawking temperature. However, using the
analogy between the Schwinger mechanism in the Minkowski spacetime
and the tunneling process in the Rindler spacetime of a black hole,
we formulate the emission rate as a contour integral in both cases.
This formulation avoids the coordinate-dependence of the emission
rate in that it does not require the black hole coordinates first to
express the emission rate and then to transform it to the Rindler
coordinates. Only the surface gravity and local Rindler coordinates
near horizons are needed for the calculation. In fact, our tunneling
(emission) rate yields a Boltzmann factor with the correct Hawking
temperature for non-extremal black holes such as a Schwarzschild
black hole, a Reissner-Nordstr\"{o}m black hole, a charged Kerr
black hole, a de Sitter space, and a Schwarzschild-anti de Sitter
black hole.

One of the reasons for using the Rindler coordinate near the event
horizon is that the tunneling process of fields in the Rindler
spacetime is analogous to the Schwinger process for pair production
by an electric field. Also the field equation in the Rindler
spacetime is quite similar to the equation minimally coupled with
the gauge field of electric field. Indeed, the tunneling (emission)
rate given by a contour integral takes the same form in the Rindler
coordinate of black holes for tunneling process and in the Minkowski
for the Schwinger mechanism. Thus the tunneling idea of fields or
particles crossing the horizon could be realized within the context
of quantum field theory in a fixed spacetime background. Our
emission rate in the Rindler coordinate has advantageous points: it
resolves the controversy of coordinate-dependence of emission rate
and the instanton action for the emission rate is indeed invariant
under canonical transformations. A caveat, however, is that our
formula cannot be applied to extremal black holes since they do not
have Rindler coordinates near horizons.

\section*{Acknowledgments}

The author would like to thank R. B. Mann, D. N. Page and other
participants of the APCTP Black Hole Workshop 2007 for useful
comments and discussions. The author also would like to appreciate
the warm hospitality of the Center for Quantum Spacetime (CQUeST),
where part of this work was done. This work was supported by the
Korea Science and Engineering Foundation (KOSEF) grant funded by the
Korea government (MOST) (No. F01-2007-000-10188-0).


\begin{thebibliography}{99}

\bibitem{Hawking} S.~W.~Hawking, ``Particle creation by black holes'',
Commun.\ Math.\ Phys. {\bf 43} (1975) 199.

\bibitem{Unruh} W.~G.~Unruh, ``Notes on black hole
evaporation,'' Phys.\ Rev.\ D {\bf 14} (1976) 870.

\bibitem{Israel} W.~Israel, ``Thermo-field dynamics of black holes,''
Phys. Lett. A {\bf 57} (1976) 107.

\bibitem{Parikh} M.~K.~Parikh and F.~Wilczek, ``Hawking radiation
as tunneling,'' Phys.\ Rev.\ Lett.\  {\bf 85} (2000) 5042
[arXiv:hep-th/9907001].

\bibitem{Schwinger} J.~Schwinger, ``On gauge invariance
and vacuum polarization,'' Phys.\ Rev.\ {\bf 82} (1951) 664.

\bibitem{Kraus95-1} P.~Kraus and F.~Wilczek,
``Self-interaction correction to black hole radiance,'' Nucl.\
Phys.\ B {\bf 433} (1995) 403 [arXiv:gr-qc/9408003].

\bibitem{Kraus95-2} P.~Kraus and F.~Wilczek,
``Effect of self-interaction on charged black hole radiance,''
Nucl.\ Phys.\ B {\bf 437} (1995) 231 [arXiv:hep-th/9411219].

\bibitem{Kraus97} P.~Kraus and E.~Keski-Vakkuri, ``Microcanonical
D-branes and back reaction,'' Nucl.\ Phys.\
B {\bf 491} (1997) 249 [arXiv:hep-th/9610045].

\bibitem{Berezin} V.~A.~Berezin, A.~M.~Boyarsky, and A.~Y.~Neronov,
``On the Mechanism of Hawking Radiation,'' Grav.\ Cosmol.\ {\bf 5}
(1999) [arXiv:gr-qc/0605099].

\bibitem{Hemming} S.~Hemming and E.~Keski-Vakkuri,
``Hawking radiation from AdS black holes,'' Phys.\ Rev.\ D {\bf 64}
(2001) 044006 [arXiv:gr-qc/0005115].

\bibitem{Vagenas01} E.~C.~Vagenas,
``Are extremal  black hole really frozen?,'' Phys.\ Lett.\ B  {\bf
503} (2001) 399  [arXiv:hep-th/0209185].

\bibitem{Vagenas02-2} E.~C.~Vagenas, ``Semiclassical corrections to
the Bekenstein-Hawking entropy of the BTZ black hole via
self-gravitation,'' Phys.\ Lett.\ B {\bf 533} (2002) 302
[arXiv:hep-th/0109108 ].

\bibitem{Medved02} A.~J.~M.~Medved, ``Radiation via tunneling in the
charged BTZ black hole,'' Class.\ Quant.\ Grav.\  {\bf 19} (2002)
589 [arXiv:hep-th/0110289].

\bibitem{Parikh02} M.~K.~Parikh, ``New Coordinates for de Sitter Space
and de Sitter Radiation,'' Phys.\ Lett.\ B {\bf 546} (2002) 189
[arXiv:hep-th/0204107].

\bibitem{Medved02-2} A.~J.~M.~Medved, ``Radiation via tunneling
from a de Sitter Cosmological Horizon,'' Phys.\ Rev.\ D {\bf 66}
(2002) 124009 [arXiv:hep-th/0207247].

\bibitem{Vagenas03} E.~C.~Vagenas, ``Generalization of the KKW analysis
for black hole radiation,'' Phys.\ Lett.\ B {\bf 559} (2003) 65
[arXiv:hep-th/0209185].

\bibitem{Padmanabhan} T.~Padmanabhan, ``Gravity and the thermodynamics
of horizons,'' Phys.\ Rept.\  {\bf 406} (2005) 49
[arXiv:gr-qc/0311036].

\bibitem{Arzano} M.~Arzano, A.~J.~M.~Medved, and E.~C.~Vagenas,
``Hawking radiation as tunneling through the quantum horizon,''
JHEP\ 0509 (2005) 037 [arXiv:hep-th/0505266].

\bibitem{Zhang} J.~Zhang and Z.~Zhao, ``Hawking radiation of charged
particles via tunneling from the Reissner-Nordstr\"{o}m black
hole,'' JHEP\ 0510 (2005) 055.

\bibitem{Zhang2} J.~Zhang and Z.~Zhao,
``Massive particles' black hole tunneling and de Sitter tunneling,''
Nucl.\ Phys.\ B {\bf 725} (2005) 173.

\bibitem{Wenbiao Liu} W.~Liu, ``New Coordinates for BTZ black
hole and Hawking radiation via tunnelling,'' Phys.\ Lett.\ B {\bf
634} (2006) 541 [arXiv:gr-qc/0512099].

\bibitem{Zhang} J.~Zhang and Z.~Zhao,
``Charged particles' tunnelling from the Kerr-Newman black hole,''
Phys.\ Lett.\ B {\bf 638} (2006) 110  [arXiv:gr-qc/0512153].

\bibitem{Jiang} Q.-Q.~Jiang, S.-Q.~Wu, and X.~Cai,
``Hawking radiation as tunneling from the Kerr and Kerr-Newman black
holes,'' Phys.\ Rev.\ D {\bf 73} (2006) 064003
[arXiv:hep-th/0512351].

\bibitem{Wu} S.-Q.~Wu, Q.-Q.~Jiang, ``Remarks on
Hawking radiation as tunneling from the BTZ black holes,'' JHEP\
0603 (2006) 079 [arXiv:hep-th/0602033].

\bibitem{Kerner06} R.~Kerner and R.~B.~Mann, ``Tunneling,
temperature, and Taub-NUT black holes,'' Phys.\ Rev.\ D {\bf 73}
(2006) 104010 [arXiv:gr-qc/0603019].

\bibitem{Wu} X.~Wu and S.~Gao, ``Tunneling effect near a weakly
isolated horizon,'' Phys.\ Rev.\ D {\bf 75} (2007) 044027
[arXiv:gr-qc/0702033].

\bibitem{Gao} L.~Gao, H.~Zhang, and W.~Liu, ``From Schwarzschild to
Kerr black hole quantum tunneling,'' Int.\ J.\ Theor.\ Phys.\ {\bf
46} (2007) 33 .

\bibitem{Kerner07} R.~Kerner and R.~B.~Mann, ``Tunneling from
G\"{o}del black holes,'' Phys.\ Rev.\ D {\bf 75} (2007) 084022
[arXiv:hep-th/0701107].

\bibitem{Srinivasan} K.~Srinivasan and T.~Padmanabhan,
``Particle production and complex path analysis,'' Phys.\ Rev.\ D
{\bf 60} (1999) 024007 [arXiv:gr-qc/9812028].

\bibitem{Shankaranarayanan01} S.~Shankaranarayanan, K.~Srinivasan,
and T.~Padmanabhan,
  ``Method of complex paths and general covariance of Hawking radiation,''
  Mod.\ Phys.\ Lett.\ A {\bf 16} (2001) 571
  [arXiv:gr-qc/0007022].

\bibitem{Shankaranarayanan02} S.~Shankaranarayanan, T.~Padmanabhan
and K.~Srinivasan, ``Hawking radiation in different coordinate
settings: complex paths approach,'' Class.\ Quant.\ Grav.\  {\bf 19}
(2002) 2671 [arXiv:gr-qc/0010042].

\bibitem{Vagenas02} E.~C.~Vagenas, ``Complex Paths and Covariance of
Hawking Radiation in 2D Stringy Black Holes,'' Nuovo\ Cimento\ B
{\bf 117} (2002) 899 [arXiv:hep-th/0111047].

\bibitem{Shankaranarayanan03} S.~Shankaranarayanan, ``Temperature and
entropy of Schwarzschild-de Sitter space-time,'' Phys.\ Rev.\ D {\bf
67} (2003) 084026 [arXiv:gr-qc/0301090].

\bibitem{Padmanabhan04} T.~Padmanabhan, ``Entropy of horizons,
complex paths and quantum tunneling,'' Mod.\ Phys.\ Lett.\ A {\bf
19} (2004) 2637  [arXiv:gr-qc/0405072].

\bibitem{Angheben} M.~Angheben, M.~Nadalini, L.~Vanzo, and S.~Zerbini,
``Hawking radiation as tunneling for extremal and rotating black
holes,''  JHEP\ 0505 (2005) 014 [arXiv:hep-th/0503081].

\bibitem{Nadalini} M.~Nadalini, L.~Vanzo, and S.~Zerbini,
``Hawking radiation as tunneling: the D-dimensional rotating case,''
J.\ Phys.\ A: Math.\ Gen. {\bf 39} (2006) 6601
[arXiv:hep-th/0511250].

\bibitem{Chowdhury} B.~D.~ Chowdhury, ``Tunneling of Thin Shells
from Black Holes: An Ill Defined Problem,'' [arXiv:hep-th/065197].

\bibitem{Akhmedov} E.~T.~Akhmedov, V.~Akhmedova, and D.~Singleton,
``Hawking temperature in the tunneling picture,''
Phys.\ Lett.\ B {\bf 642} (2006) 124 [arXiv:hep-th/0608098].

\bibitem{Akhmedov07} E.~T.~Akhmedov, V.~Akhmedova, D.~Singleton, and
T.~Pilling, ``Thermal radiation of various gravitational
backgrounds,''  Int.\ J.\ Mod.\ Phys.\ A {\bf 22} (2007) 1705
[arXiv:hep-th/0605137].

\bibitem{Nakamura} T.~K.~Nakamura, ``Factor of Two Discrepancy of
Hawking Radiation Temperature,'' [arXiv:0706.2916].

\bibitem{Kim07} S.~P.~Kim, ``Schwinger mechanism and Hawking
radiation as quantum tunneling,'' [arXiv:0709.4313].

\bibitem{Brout} R.~Brout, R.~Parentani, and Ph.~Spindel,
``Thermal properties of pairs produced by an
electric field: A tunneling approach,'' Nucl.\ Phys.\ B {\bf 353}
(1991) 209.

\bibitem{Parentani92} R.~Parentani and R.~Brout,
``Vacuum instability and black hole evaporation,''
 Nucl.\ Phys.\ B {\bf 388} (1992) 474.

 \bibitem{BMPS} R.~Brout, S.~Massar, R.~Parentani, and Ph.~Spindel,
``A primer for black hole quantum physics,'' Phys.\ Rep.\ {\bf 260}
(1995) 329.

\bibitem{Parentani97} R.~Parentani and S.~Massar, ``Schwinger
mechanism, Unruh effect, and production of accelerated black
holes,'' Phys.\ Rev.\ D {\bf 55} (1997) 3603 [arXiv:hep-th/9603057].

\bibitem{Kim-Page07} S.~P.~Kim and D.~N.~Page, ``Improved
approximations for fermion pair production in inhomogeneous electric
fields,'' Phys.\ Rev.\ D {\bf 75} (2007) 045013
[arXiv:hep-th/0701047].

\bibitem{Kim-Page02} S.~P.~Kim and D.~N.~Page, ``Schwinger
pair production via instantons in a strong electric field,'' Phys.\
Rev.\ D {\bf 65} (2002) 105002 [arXiv:hep-th/0005078].

\bibitem{Kim-Page06} S.~P.~Kim and D.~N.~Page,
``Schwinger pair production in electric and magnetic fields,''
Phys.\ Rev.\ D {\bf 73} (2006) 065020 [arXiv:hep-th/0301132].

\end{thebibliography}
\end{document}